# Feasibility study of heavy ion beams and compound target materials for muon production

**Jaebum Son, Ju Hahn Lee[*], Gi Dong Kim, Yong Kyun Kim**

*Institute for Basic Science, Daejeon 305-811*

We have investigated the feasibility of using compound materials as target for muon production by virtue of simulations using a GEANT4 toolkit. A graphite and two thermostable compound materials, beryllium oxide (BeO) and boron carbide ($B_4C$) were considered as muon production targets and their muon production rates for 600-MeV proton beam were calculated and compared. For thermal analysis, total heat deposited on the targets by the proton beams and the secondary particles was calculated with a MCNPX code, and then the temperature distribution of target was derived from the calculated heat by using an ANSYS code with consideration for heat transfer mechanisms, such as thermal conduction and thermal radiation. In addition, we have investigated whether the heavy ion beams can be utilized for muon production. For various beam species such as $^3He^2$, $^4He$, $^7Li$, $^{10}B$ and $^{12}C$, their muon production rates were calculated and compared with that obtained for a proton beam.

PACS number: 24.10.Lx

Keywords: Muon, Muon Spin Rotation/Relaxation/Resonance, Muon production target, RISP

# I. INTRODUCTION

Muon spin rotation/relaxation/resonance (μSR) is regarded one of the most powerful tool in material science due to its high sensitivity for weak and dynamic internal fields [1]. Three decades after the advent of μSR, only four μSR facilities have been successfully operating in the world [2-5]. Fortunately, several facilities are expected to be constructed in the near future, and one of them is currently underway at South Korea [6]. The development of the RISP μSR facility is starting from only an exclusive muon beam line for μSR. However it will be gradually extended for conducting several μSR experiments and muon science at the same time. Considering the extension, we are currently designing a target system for obtaining the highest surface muon yield with the accelerator complex of RISP, RAON.

Since the main driver of RAON can deliver various beams like proton to uranium with a beam power of 400 kW [7], various combinations of primary beam species and target material may be considered for muon production. Several restrictions are imposed on primary beam and target material as these will contribute to decide muon production rate. Among them, the primary beam energy is the most important one because it has to exceed a threshold level (~289 MeV per nucleon) to produce pions that decay to muons. Considering the maximum beam energy capable of being reached from the RAON driver, only $^3$He, $^4$He, $^7$Li, $^{10}$B and $^{12}$C as well as proton can be taken into account as primary beam for muon production.

In choosing a target material for muon production, melting point, thermal conductivity and safety hazard in handling are as important as muon production rate. Although it depends on the path length of incident beam inside a target, the energy deposited on muon production target generally reaches at least 10% of incident beam energy. Furthermore, since the energy is totally accumulated on the target as heat, the material must have the strong characteristics in heat. For this reason, graphite and beryllium have been mainly used for muon production until now. Recently a lot of effort has been put into developing high-power target even for radioactive beam production, and various materials like liquid lithium (LLi), beryllium oxide (BeO) and boron carbide ($B_4C$) has been investigated as target material. Considering their good thermal characteristics and easy handling, they are worth being seriously concerned for muon production.

In the present work, to compare muon production yields for several beam species and target materials, Monte Carlo simulations were performed for using a GEANT4 toolkit [8]. In addition, to obtain total heat deposited on the target by the primary beams and the secondary particles was calculated with a MCNPX code [9], and then the temperature distribution of target was derived from the calculated heat by using an ANSYS code [10] with consideration for heat transfer mechanisms, such as thermal conduction and thermal radiation.

## II. SIMULATIONS AND DISCUSSION

Simulations were performed using a Bertini model which is a physic model commonly used for applying the quark gluon model for high energy interactions of protons, neutrons, pions and kaons below ~10 GeV physics. The target was set to have a cylindrical shape with a diameter of 3 mm and with a length of 6 cm. Primary beam was considered as continuous beam with a Gaussian shape of $1\sigma = 1$ mm. Actually the diameter of target and beam spot size are important factors to decide a production yield for muon, especially for surface muon. Our previous work, we found that the production yield of surface muon has the maximum value when the width of target is comparable to beam spot size (see Fig. 1). The primary beam was set to be incident on the center of circular area as perpendicular to the surface of circular area. As mentioned above, $^3$He, $^4$He, $^7$Li, $^{10}$B and $^{12}$C as well as proton were used as primary beam. Considering the angular distribution as well as the production yield of muon, virtual detector was designed as spherical shell with a diameter of 50 cm, which consisted of 18 ring-type segments. Each segment has an angular coverage of 10 degrees. The target was located at the center of virtual detector and the beam generator was positioned at 10 cm in front of the circular surface of the target inside the virtual detector (see Fig. 2). Total $10^9$ beam particles were generated for each condition.

In order to check the validation of our simulation code, we looked into the energy spectrum of positive muons and the angular distribution of surface muon. Fig. 3 shows energy spectra of positive muons for various proton energies. The peaks corresponding to surface muon which has a kinetic energy of ~ 4.1 MeV and their tails due to an energy loss of surface muon in the target are clearly shown in Fig. 3 and the inset of Fig. 3. It is reasonable that the peak position does not depend on beam energy but the production yield increases as the beam energy increases. Considering the target geometry and a production mechanism of surface muon, the surface muons should be anisotropically emitted from the target, which is clearly identified in the angular distribution plot for surface muon (see Fig 4). This anisotropic emission of surface muon can be seriously considered to design the extraction port of surface muon.

We performed the simulations to obtain muon production yields for various beam species. The beam energy was 400 MeV/nucleon for other species except proton, which is available at RAON. The graphite was utilized as target material. Fig. 5 shows the number of muons produced for total $10^9$ of beam particles. Other beams except $^3$He produced slightly more negative muons but positive muons much less than those by proton beam. Especially the number of surface muons produced by other beams was about 10 times smaller than that by proton beam. Therefore if one has to use the surface muon the proton might be used as the primary beam. However, if the negative moun is required, lithium or boron may be substituted for the proton beam.

For target material, we conducted the simulations with a 600-MeV proton for graphite, LLi, BeO, and $B_4C$. These target materials were chosen by considering their melting points and thermal conductivity (see Fig. 6). The numbers of muons obtained by simulations for these materials are shown in Fig. 7. Except LLi, although they seem to have broadly similar muon production rates for each other, it is certain that the muon production rate of $B_4C$ is slightly larger than those of other materials. To carry out thermal analysis for muon production targets, we assumed that the target has a disk shape with an inner hallow and circumrotates with a speed of 60 rev/min. The disk-like target has an outer diameter of 30 cm, an inner diameter of 20 cm and a thickness of 3 mm (see Fig. 8(a)). The primary beams are supposed to be incident on the side of the disk-like target and to travel 20-cm distance inside the target. By using the MCNPX code, the distribution of heat deposited on the disk-like target by an irradiation of only one particle was calculated (see Fig. 8(b)), and then the total thermal power distribution on the target was derived in consideration of the rotation speed of target and beam irradiation conditions such as a CW beam mode and a beam power of 400 kW. Regarding this calculated thermal power distribution as an input parameter, the temperature distribution on the target was calculated by using the ANSYS code (see Fig. 8(c)). In the calculations, we only considered heat transfer mechanisms such as thermal conduction and thermal radiation, and therefore used the thermal conductivities of target materials shown in Fig. 6 and the emissivity value of 0.8 for all materials. The maximum and minimum temperatures of graphite and boron carbide targets for proton and boron beams are listed in Table 1. For all cases, the maximum temperature of target does not exceed the melting point of target material. Therefore, at least all combinations of target materials and beam species in Table 1 are expected to be used for muon production.

### III. CONCLUSION

In conclusion, we investigated the suitable target material and beam species for muon production. Muon production yields were obtained by using a GEANT4 toolkit and for the thermal analysis of a disk-shape rotational target, total deposited heat on target was calculated by using a MCNPX code, and the temperature distribution of target was derived from the calculated heat distribution by using an ANSYS code. As results, according to the kind of muon required, lithium or boron could be substituted for proton as a primary beam species. Whereas, in case of target material, boron carbide showed better performance on muon production than graphite and enough thermal characteristics to be used as muon production target. Therefore, considering these results, the boron carbide target could replace the graphite target that is now widely used for muon production. In the future, more detailed thermal analysis and mechanical analysis such as deformation and cracking due to thermal stress will be carried out for designing the realistic target structure.

## ACKNOWLEDGEMENT

This work was supported by the Rare Isotope Science Project of the Institute for Basic Science funded by the Ministry of Science, ICT, and Future Planning and National Research Foundation of Korea (2014M7A1A1075765).

## REREFENCES

Table 1. The maximum and minimum temperatures of graphite and boron carbide targets for the proton and boron beams with 400 kW.

|  | 600-MeV proton beam | | 400-MeV/nucleon boron beam | |
|---|---|---|---|---|
|  | Graphite (M.P. 3,652 K) | $B_4C$ (M.P. 3,036 K) | Graphite | $B_4C$ |
| Max. Temp.(K) | 2,431 | 2,500 | 2,338 | 2,362 |
| Min. Temp.(K) | 1,633 | 1,740 | 1,652 | 1,559 |

**Figure Captions.**

Fig. 1. Dependence of surface muon production yield on a width of target [11].

Fig. 2. Simulation conditions including a physics model, a target and virtual detector geometries and beam species used in this work.

Fig. 3. Energy spectra of positive muons for various proton beam energies. The inset shows the peaks corresponding to surface muons.

Fig. 4. Angular distribution for surface muons detected on the virtual detector.

Fig. 5. The number of muons obtained with a graphite target by simulations for $^{3}$He, $^{4}$He, $^{7}$Li, $^{10}$B and $^{12}$C as well as proton beams.

Fig. 6. Thermal conductivities and melting points for graphite, liquid lithium (LLi), beryllium oxide (BeO), and boron carbide ($B_4C$).

Fig. 7. The number of muons obtained with a 600-MeV proton beam by simulations for various target materials, such as graphite, LLi, BeO and $B_4C$.

Fig. 8. (a) Illustration of a disk-like rotation target designed for thermal analysis, (b) the calculated distribution of heat deposited on the disk-like target by an irradiation of only one particle by a MCNPX code and (c) the temperature distribution on the target calculated by an ANSYS code.

## Simulation conditions

- Target material: Graphite (density: 1.8 g/cm$^3$, melting point: 3300 K)
- Primary beam: Proton (600 MeV), Gaussian type (1$\sigma$ ~ 0.5), height: 5 cm, length: 6 cm
- Muon detector: Plastic detector (diameter: 50 cm, thickness: 6 cm)

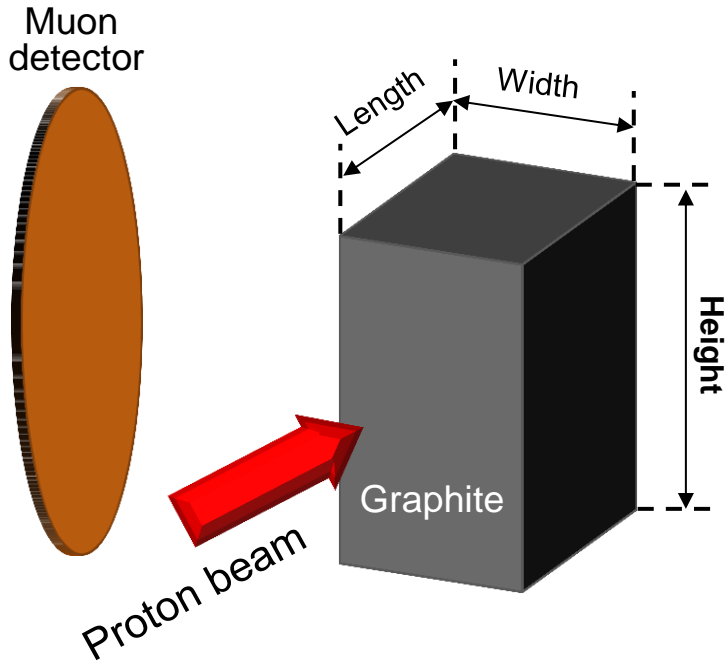
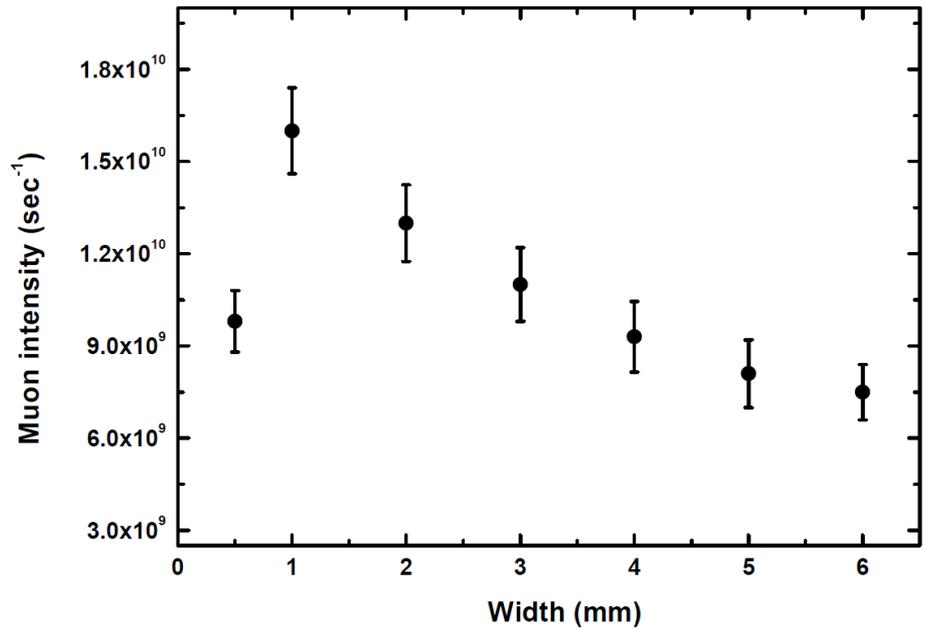

- GEANT4 toolkit
- Bertini model (QGSP_BERT)
- Target: Cylindrical type ($\phi$ 3 mm, Length 6 cm), graphite, BeO, $B_4C$
- Primary beam: Gaussian shape (1$\sigma$~1 mm), $H^+$, $^4He^{2+}$, $^7Li^{3+}$, $^{10}B^{5+}$, $^{12}C^{6+}$
- Detector: Spherical shell type, 18 ring segments
    (10 degrees of angular coverage for each segment)

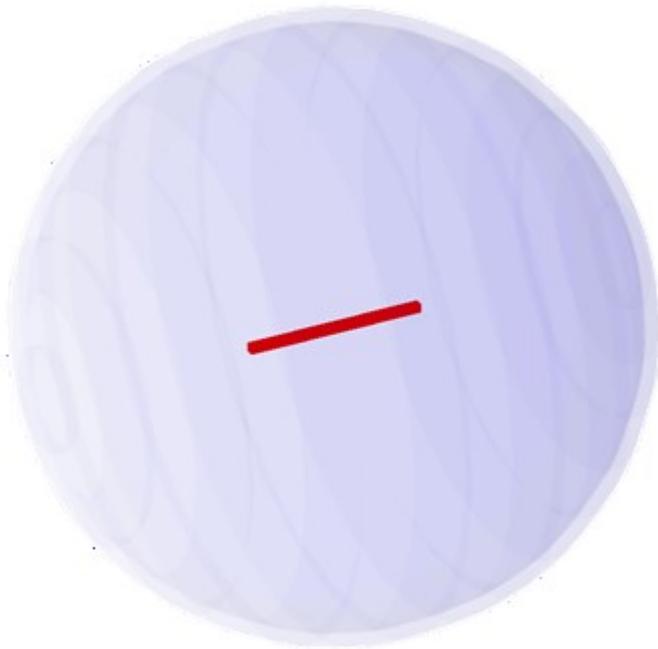
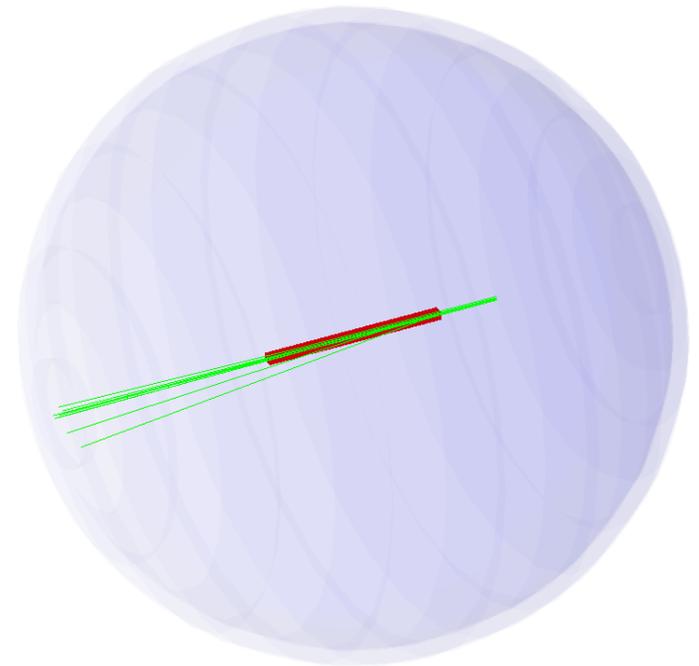

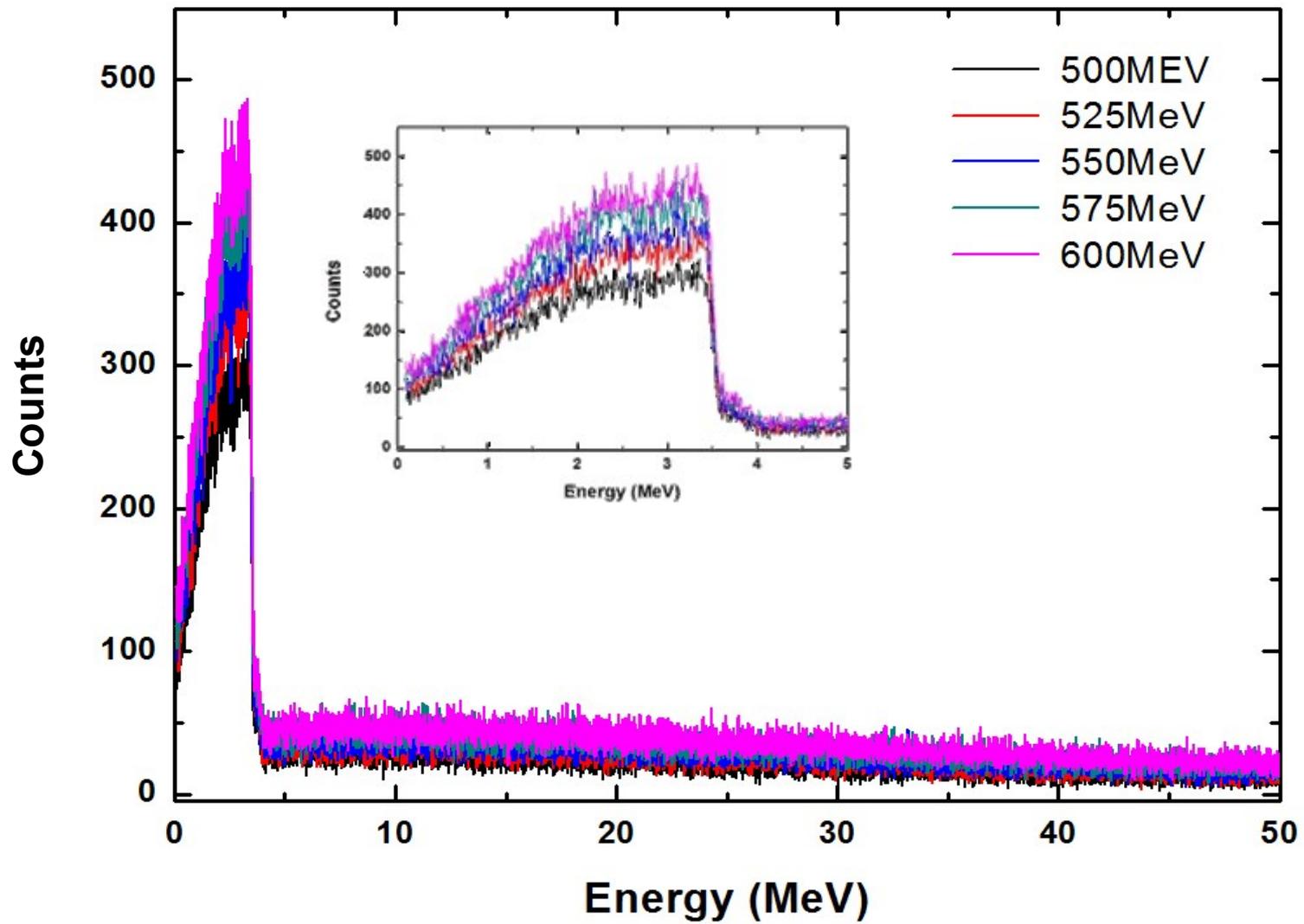

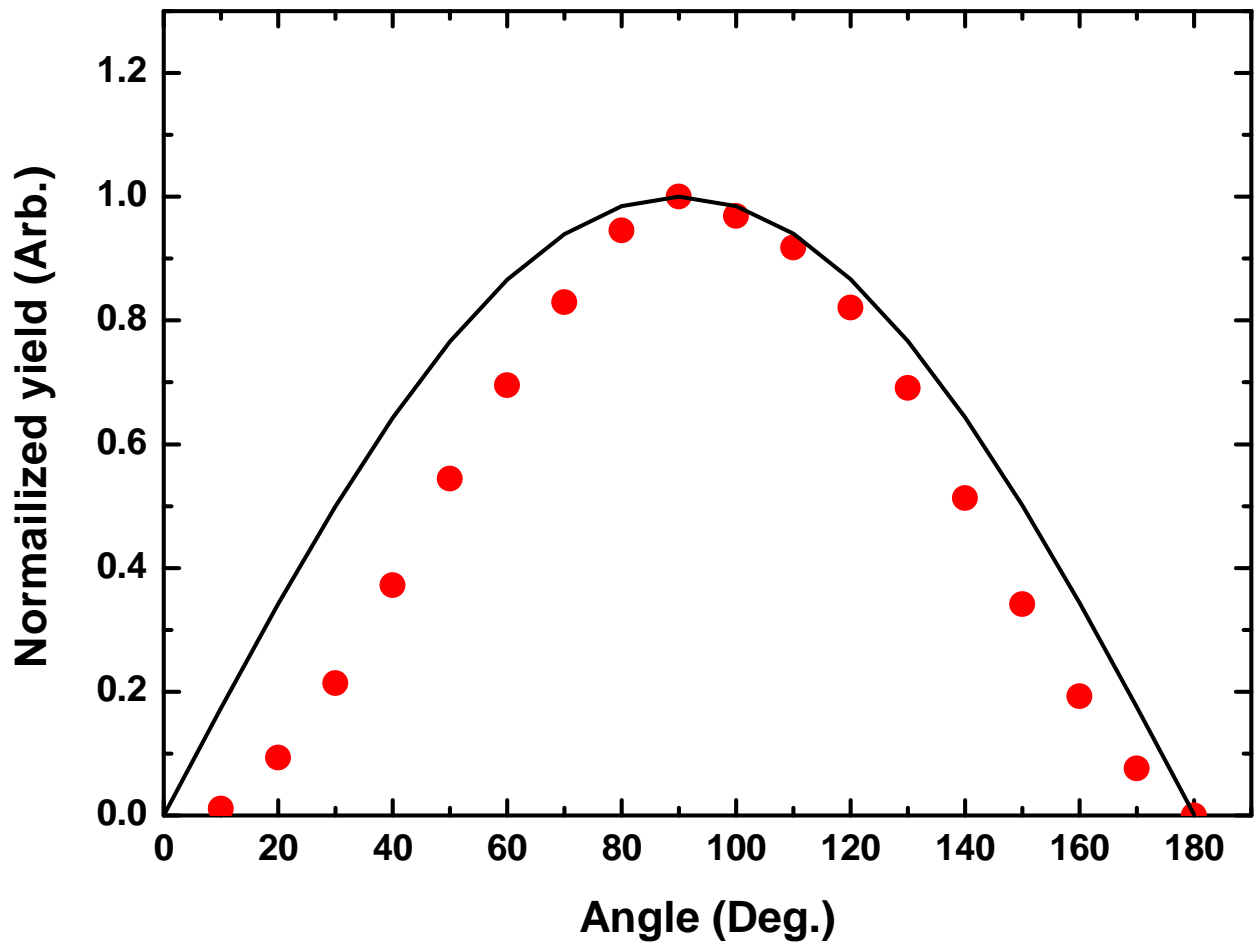

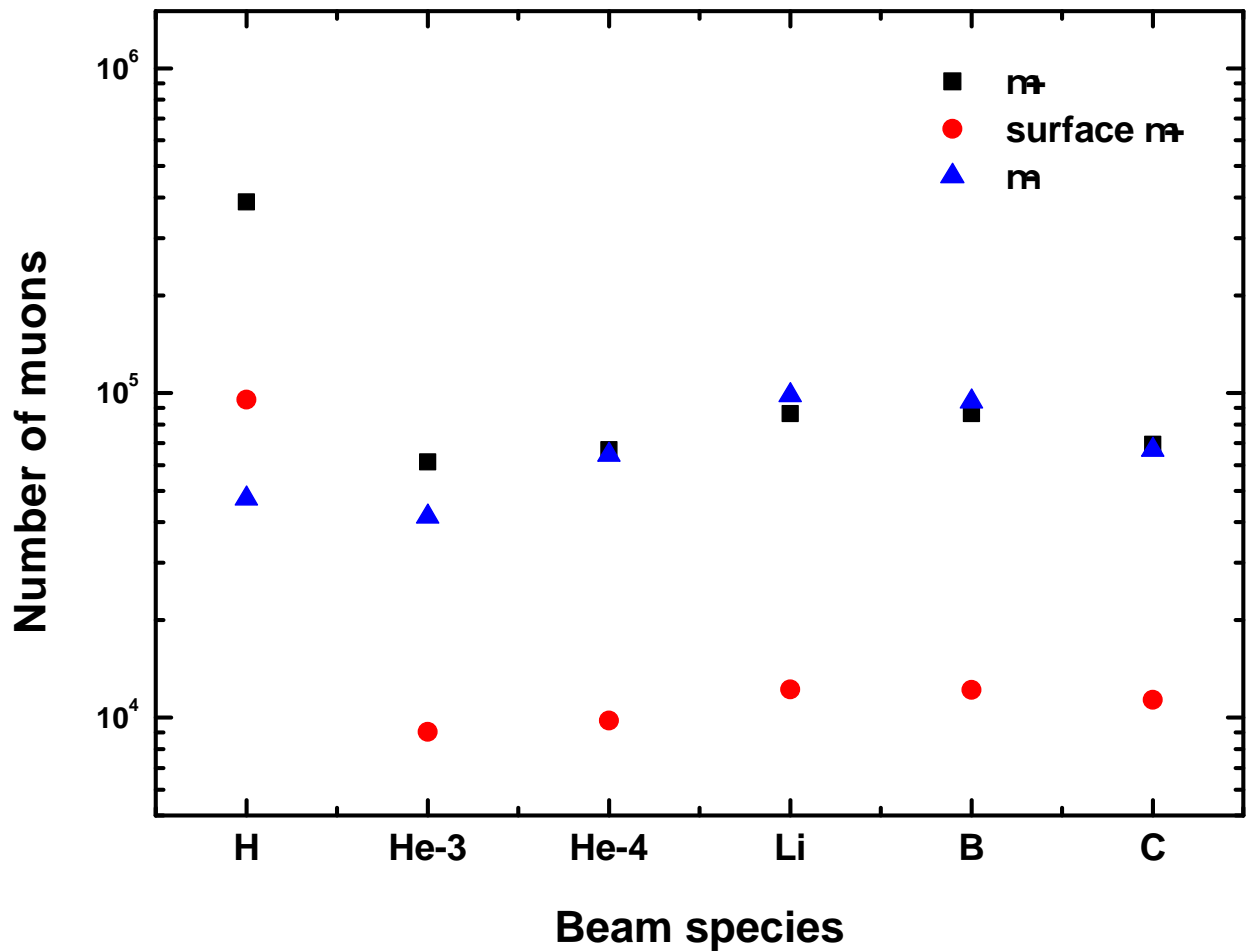

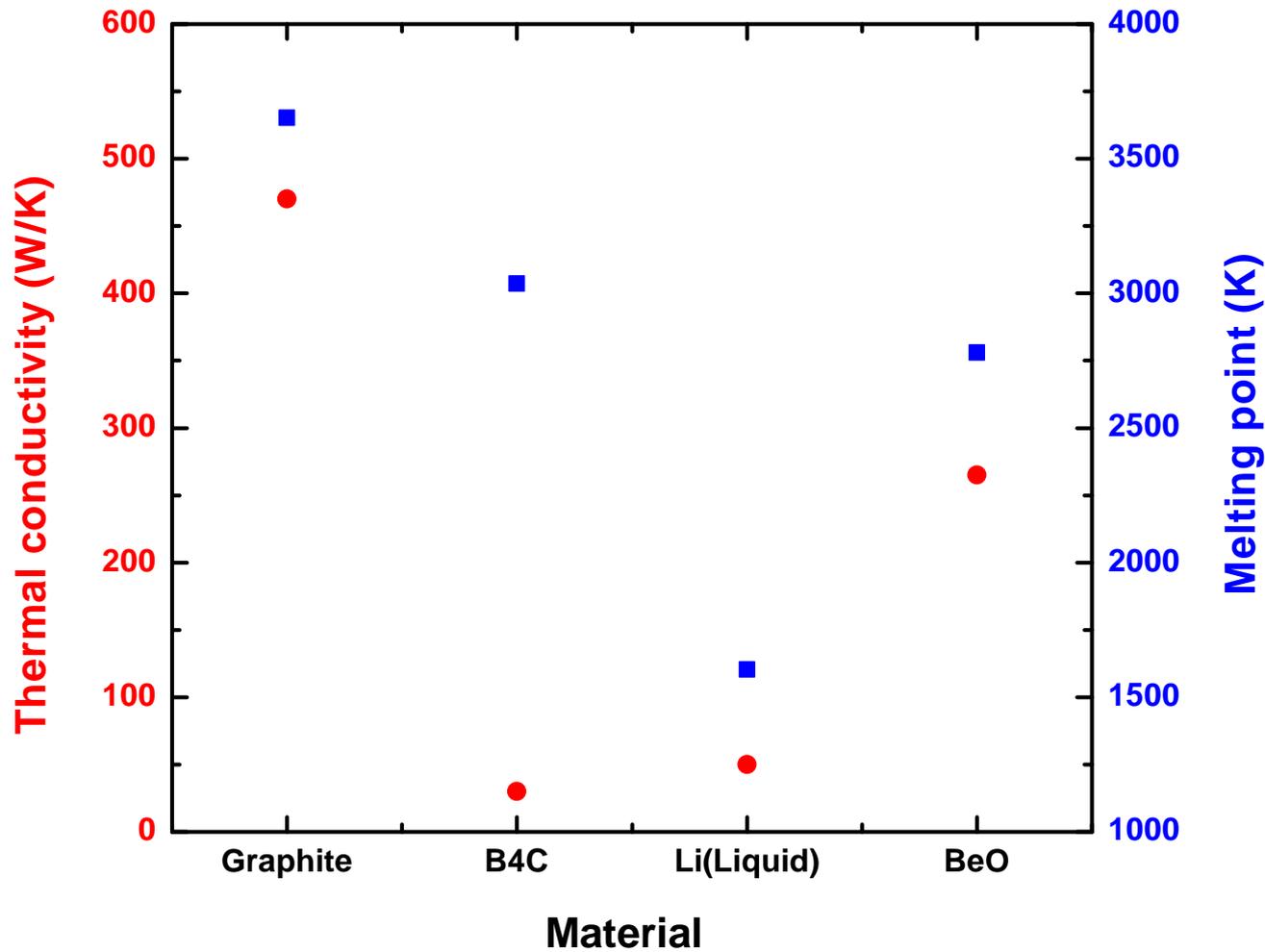

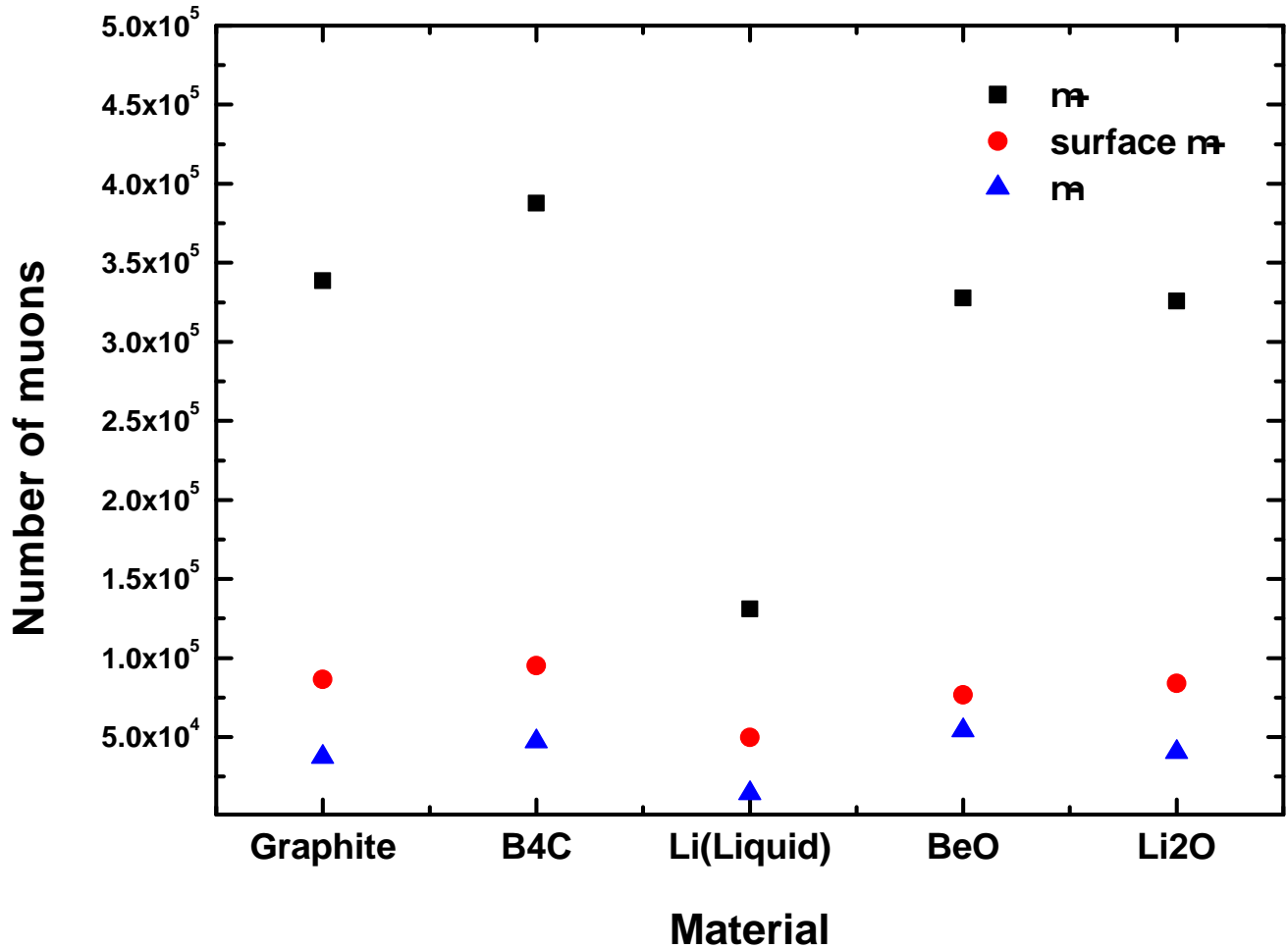

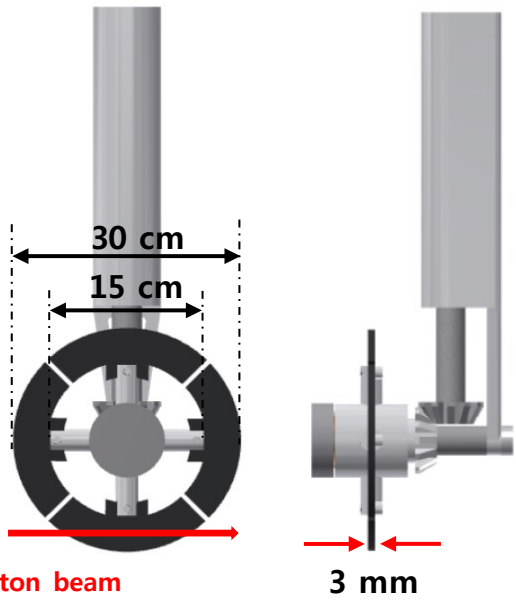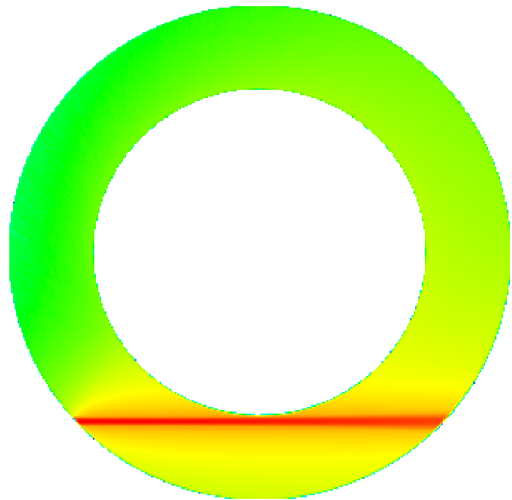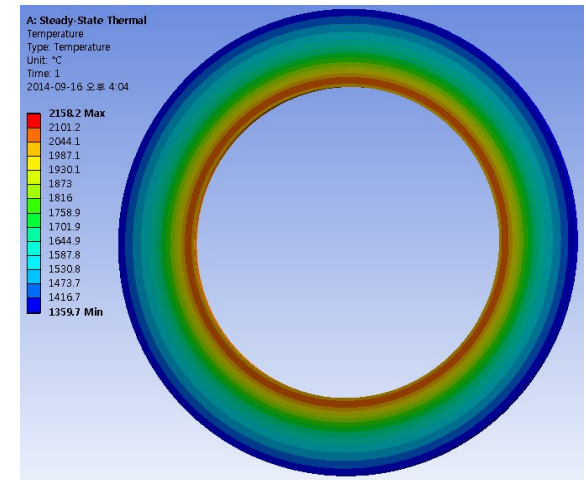